\documentclass[twocolumn, 11pt, a4paper, floatfix, 10pt]{revtex4-2}
\usepackage{amsmath, amssymb, bm, graphicx}
\usepackage[colorlinks,allcolors=blue]{hyperref}

\newcommand{\vect}[1]{\bm{#1}}
\newcommand{\tens}[1]{\mathsf{#1}}

\begin{document}

\title{Growth phases of an active tissue: determinate, indeterminate, and proportionate}

\author{Jigyasa Watwani}
\author{K Vijay Kumar}
\author{Vishal Vasan}
\affiliation{International Centre for Theoretical Sciences, Tata Institute of Fundamental Research, \\ 
Survey 151, Shivakote Village, Hesaraghatta Hobli, 
Bengaluru North, India 560089.}

\date{\today}

\begin{abstract}
Growth may cease at a target size or continue throughout life: the determinate and indeterminate phenotypes. We develop an active viscoelastic continuum model of a tissue growing along one axis, in which cell division and death generate active stresses. We find two asymptotic states: one in which the tissue reaches a relative size fixed by its material parameters, and one in which it elongates linearly without bound. Which state is realised is set by the ratio of active stress to elastic modulus. The transition originates in a bound on the elastic stress the tissue can support: a sufficiently large activity can never be balanced. In a tissue made of parts with different material properties, the growing phase settles into fixed length proportions, set by the mechanical impedances of the parts rather than inherited; matching impedances to initial lengths preserves the proportions the tissue began with. Determinate, indeterminate and proportionate growth thus appear as regimes of one continuum mechanical framework.
\end{abstract}

\maketitle

\section{Introduction} In many organisms growth ceases at a ``target size'' and the growth curve asymptotes, as it does in birds, mammals and most vertebrates \cite{Parks1982book}. Others have no upper size limit and grow throughout life, either at a rate that decreases with age without ever reaching an asymptote, as in some crustaceans \cite{hartnoll1983}, or linearly, as in planaria \cite{Oviedo2003} and modular organisms such as annelids \cite{Hariharan2015}. These two phenotypes are called \textit{determinate} and \textit{indeterminate} growth, and the distinction extends to how each responds to perturbation. Determinate systems regulate tightly: cell sizes and numbers compensate for one another so that the final size is maintained \cite{Neufeld1998, Hisanaga2015Compensation, Fankhauser1945_salamander_compensation}, and catch-up growth in one organ after an insult to another preserves body proportions \cite{Rosello2018_crosstalk_limbs}. Indeterminate systems instead track their conditions throughout life, and to widely varying degrees: planaria span three orders of magnitude and even ``de-grow'', shrinking when starved \cite{RinkDev2019, baguna1981}. Indeterminate growth is widespread enough across the tree of life that it has been argued to be the ancestral condition, from which determinate growth was derived repeatedly \cite{Hariharan2015}; if so, the two are more likely to be separated by a change of parameter than by a change of mechanism. Whether the two phenotypes call for different mechanisms, or are instead two regimes of a single physical description, is the first of the questions we address.

A second question concerns not the size of a tissue but its proportions. Many organisms grow while holding the relative sizes of their parts nearly fixed, and restore them after a perturbation: catch-up growth preserves limb proportions \cite{Rosello2018_crosstalk_limbs}, and planaria rescale their body plan continuously as they grow and de-grow \cite{Oviedo2003, RinkDev2019}. This is \textit{proportionate} growth, and it is not obvious that a purely mechanical model should produce it, since the parts of a heterogeneous tissue have different material properties and would generically elongate at different rates.

Models of growth control are typically built on morphogens: mis-expression of \textit{dpp} in \textit{Drosophila} produces drastically different wing sizes \cite{LawrenceDay2000}. Most leave out mechanics, though tissues are known to grow in response to mechanical cues \cite{Aragona2020}. A closely related feedback couples morphogen and growth in both directions \cite{AguilarHidalgo2018}: a temporal growth rule sets the local growth rate from the relative rate of change of the morphogen concentration, while growth in turn dilutes and advects the morphogen. The growth rule is imposed rather than derived, but the feedback it closes has a critical point at which growth is spatially homogeneous and the gradient scales with tissue length, and which, according to how the degradation rate scales with size, separates growth arrest from unbounded growth. Mechanisms of the same family account for the rescaling of the planarian body plan \cite{Werner2015, Stuckemann2017}. Models that do couple mechanics to signalling \cite{Aegerter-Wilmsen2007-, Hufnagel2007, Shraiman2005-} prescribe growth by hand, and only recently has the role of mechanical fields in setting growth kinetics and termination been explored \cite{Heinrich2020eLife}.

Here we ask whether determinate and indeterminate growth can be two phases of a single mechanical model, and whether that same model can account for proportionate growth in an inhomogeneous tissue. We write a minimal coarse-grained model in which growth is implicit rather than prescribed, neglecting growth anisotropy and signalling morphogens, and find that the two phenotypes are selected by the ratio of the active stress to the tissue elastic modulus. The transition we find is of the same qualitative type as that of \cite{AguilarHidalgo2018}, bounded against unbounded growth with proportions held fixed in the growing phase, but arises from mechanics alone, with no chemical field and with no growth rate prescribed. When the tissue is driven by a stress applied at its boundary the model is solvable in closed form. When the driving is generated internally, by a cell density that itself evolves, closed-form results survive for the steady state at any turnover rate and for the growing state at high turnover; the intervening regime we treat numerically.

\begin{figure}[t]
\includegraphics[width=\linewidth]{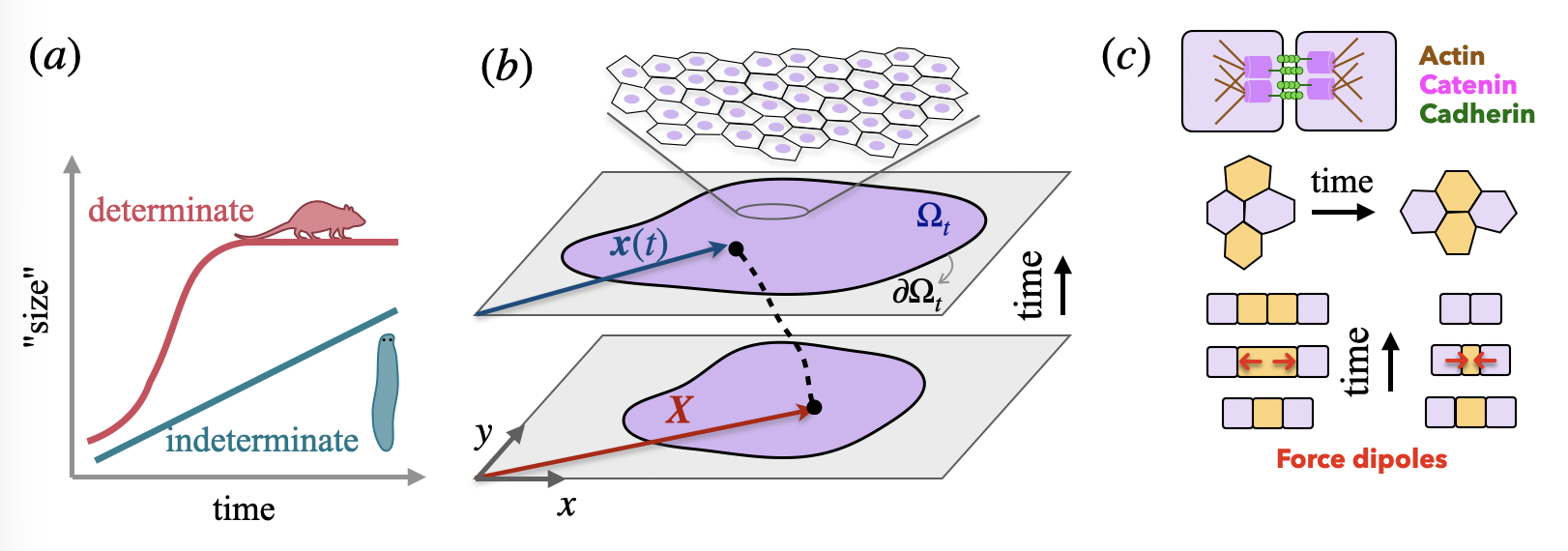}
\caption{(a) Schematic growth curves. Determinate growth saturates at a target size, as in a mouse \cite{Parks1982book}; indeterminate growth continues throughout life, here linearly, as in a planarian \cite{Oviedo2003}. (b) The tissue as an active deformable material. A material point at $\vect{X}$ in the initial configuration is carried along its trajectory (dashed) to $\vect{x}(t)$ in the domain $\Omega_t$ occupied at time $t$, with boundary $\partial\Omega_t$; inset, the cellular scale coarse-grained over. (c) Microscopic origin of the three contributions to the stress: cadherin--catenin junctions coupled to the actin cytoskeleton hold neighbouring cells together (elastic); topological rearrangements, in which cells exchange neighbours, let them flow past one another (viscous); and division and extrusion act as extensile and contractile force dipoles (active).}
\label{fig:schematic}
\end{figure}

\section{Continuum model of a tissue} We model a growing tissue as a viscoelastic continuum of finite size with a boundary (Fig.~\ref{fig:schematic}(b)). Material points in the tissue are labelled by their position vectors $\vect{x}(t)$, and the collection of all such points defines the time-dependent domain $\Omega_t$ and its boundary $\partial\Omega_t$. The tissue deforms due to a velocity field $\vect{v}(t, \vect{x})$ defined at each point $\vect{x} \in \Omega_t$. Clearly,
\begin{align}
\dot{\vect{x}} = \vect{v}(t, \vect{x}),
\label{eq:material_point_evolution}
\end{align}
where an overdot denotes a time derivative taken following a material point. At $t=0$ we set $\vect{x}(0)=\vect{X}$; the set of all $\vect{X}$ defines the initial configuration $\Omega_0$. We denote by $\vect{x}(t, \vect{X})$ the solution to \eqref{eq:material_point_evolution} starting from the initial condition $\vect{X}$. The displacement field is defined by $\vect{u}(t, \vect{X}) = \vect{x}(t,\vect{X}) - \vect{X}$. Provided $\nabla \cdot \vect{v}$ is bounded on $\Omega_t$, the deformation gradient $\tens{J} \equiv \partial\vect{x}/\partial\vect{X}$ is non-singular at any finite time, the map $\vect{X} \mapsto \vect{x}$ is invertible, and the displacement may be regarded as a field $\vect{u}(t,\vect{x})$ on the current domain \cite{SI}. It then obeys
\begin{align}
D_t \vect{u} \equiv \left( \partial_t + \vect{v}(t, \vect{x}) \cdot \nabla \right) \vect{u}(t, \vect{x}) = \vect{v}(t, \vect{x}),
\label{eq:displacement_evolution}
\end{align}
where the gradient $\nabla$ is computed with respect to $\vect{x}$. Equations \eqref{eq:material_point_evolution} and \eqref{eq:displacement_evolution} determine the evolution of both the domain $\Omega_t$ and the displacement field defined on it, once $\vect{v}(t,\vect{x})$ is specified. Sought within the class of affine displacements $\vect{u} = \tens{A}\cdot\vect{x}$, with $\tens{A}$ a constant tensor, \eqref{eq:displacement_evolution} admits for an isotropic tissue exactly two asymptotic states of this form \cite{SI}: $\vect{v} = \vect{0}$, in which $\Omega_t$ is static, and $\vect{u} = \vect{x}$, in which it deforms perennially. The second is precisely the configuration in which the Lagrangian map ceases to be invertible, since $\vect{u} = \vect{x}$ means $\nabla\vect{u} = \tens{I}$, at which $\tens{J} = (\tens{I}-\nabla\vect{u})^{-1}$ is singular and $\vect{X} = \vect{0}$. The perennially deforming state is thus indeterminate in two ways -- the domain grows without bound, and all memory of $\Omega_0$ is lost. It is approached only asymptotically \cite{SI}, and we return to it in the Discussion.

The description thus far is entirely kinematical; to close the system we take the velocity at time $t$ to be determined by the instantaneous domain $\Omega_t$ together with the displacement field $\vect{u}(t,\vect{x})$ defined on it. Since $\vect{u}$ is the integral of $\vect{v}$ along material trajectories, this retains the full deformation history relative to $\Omega_0$; what it excludes is any \emph{additional} memory, such as a relaxation kernel acting on the strain-rate history. The exclusion is a genuine restriction for a growing tissue, since division and apoptosis themselves relieve stress: at rate $\kappa$ they fluidise the tissue on times long compared with $\kappa^{-1}$ \cite{Ranft2010}. Our fixed reference configuration is the complementary limit, in which that relaxation is slow compared with the time over which the tissue grows.

At the scale of tissues, dynamics is typically overdamped, so that inertia is negligible. Although the tissue does not conserve mass, cell division and death events carry no net momentum at the coarse-grained level, and so momentum balance reduces, at each instant of time, to a quasi-static force-balance equation:
\begin{align}
\nabla \cdot \tens{\sigma} + \vect{F}_{\mathrm{ext}} = 0,
\label{eq:force_balance}
\end{align}
where $\vect{F}_{\mathrm{ext}}$ is any external force and $\tens{\sigma}$ is the Cauchy stress, the sum of elastic, viscous and active contributions, $\tens{\sigma} = \tens{\sigma}_{\mathrm{elastic}} + \tens{\sigma}_{\mathrm{viscous}} + \tens{\sigma}_{\mathrm{active}}$, whose microscopic origins are sketched in Fig.~\ref{fig:schematic}(c).

As remarked earlier, we assume that the mechanical stress is a function of the current configuration. The elastic and viscous stresses then arise from the instantaneous gradients of the displacement and velocity fields respectively. Writing $\tens{\varepsilon} = \tfrac{1}{2}[\nabla\vect{u} + (\nabla\vect{u})^{\mathsf{T}}]$ and $\dot{\tens{\varepsilon}} = \tfrac{1}{2}[\nabla\vect{v} + (\nabla\vect{v})^{\mathsf{T}}]$, a minimal description that captures this local behaviour is to take $\tens{\sigma}_{\mathrm{elastic}} = 2K ( \tens{\varepsilon} - \tfrac{1}{d} (\mathrm{tr}\,\tens{\varepsilon}) \tens{I} ) + E \, (\mathrm{tr}\,\tens{\varepsilon}) \, \tens{I}$ and $\tens{\sigma}_{\mathrm{viscous}} = 2\mu ( \dot{\tens{\varepsilon}} - \tfrac{1}{d} (\mathrm{tr}\,\dot{\tens{\varepsilon}}) \tens{I} ) + \eta \, (\mathrm{tr}\,\dot{\tens{\varepsilon}}) \, \tens{I}$, where $K$ and $E$ are the shear and bulk elastic moduli, and $\mu$ and $\eta$ are the shear and bulk viscosities. The active stress $\tens{\sigma}_{\mathrm{active}}$ can be locally regulated by the cell number density $\rho$, introduced explicitly in equation \eqref{eq:density_evolution} below, by local anisotropies in the tissue, as well as by signalling fields. In this study, we neglect the local anisotropic degrees of freedom of the tissue as well as any signalling fields. Specific choices for $\tens{\sigma}_{\mathrm{active}}$ will be discussed later.

The external force arises from whatever constrains the motion of the tissue. We take $\vect{F}_{\mathrm{ext}} = -\gamma \, \vect{v}$, the friction exerted by a substrate on a quasi two-dimensional tissue growing upon it. What matters below is not the particular form but that some dissipative coupling be present, so as to render the velocity in \eqref{eq:force_balance} unique; the scales it introduces, the hydrodynamic length $\ell=\sqrt{\eta/\gamma}$ and the impedance $Z=\sqrt{\eta\gamma}$, then organise all of the results that follow.

On a deforming domain the appropriate boundary condition is to specify the total mechanical stress on the periphery, $(\vect{\hat{n}} \cdot \tens{\sigma})\vert_{\partial \Omega_t} = \vect{\sigma}_b$, where the traction $\vect{\sigma}_b$ is a specified vector-valued function on $\partial\Omega_t$ and $\vect{\hat{n}}$ is the unit outward normal. Given an initial condition $\vect{u}(0,\vect{x})$ and a specified active stress $\tens{\sigma}_{\mathrm{active}}$, equations \eqref{eq:material_point_evolution}--\eqref{eq:force_balance} together with this boundary condition completely specify the problem.

Two assumptions deserve emphasis. First, the identity of the tissue is frozen: the material parameters and the form of $\tens{\sigma}_\mathrm{active}$ are fixed in time and do not respond to how much the tissue has grown, so we do not treat the coupled dynamics of growth and morphogen fields. Second, until the density field is introduced in \eqref{eq:density_evolution}, the tissue has no notion of a local cell number, and a perennially growing tissue therefore rarefies.

\section{Bounded, unbounded, and proportionate growth in response to an applied stress} We now specialise to growth along a single axis, which we take to be $x$, so that the tissue occupies $x \in \left(-\tfrac{L(t)}{2}, \tfrac{L(t)}{2} \right)$ and its transverse dimensions are either fixed or grow in proportion to $L(t)$. This describes tissues that elongate predominantly along one direction, such as the anteroposterior axis of a planarian \cite{Oviedo2003} or the proximodistal axis of a developing limb \cite{Rosello2018_crosstalk_limbs}. The governing equations are
\begin{align}
\dot{x} = v,
\quad
D_t u = v,
\quad
\partial_x \sigma = \gamma v,
\quad
\sigma \big\vert_{x = \pm \frac{L(t)}{2}} = \sigma_b,
\label{eq:1D_eqns}
\end{align}
where the total stress is $\sigma = E \partial_x u + \eta \partial_x v$, the active stress being set to zero for the present. The tissue is thus held under a uniform tension $\sigma_b$: the traction $\vect{\hat{n}} \cdot \tens{\sigma}$ points outward at both edges, so that the stress itself takes the same value $\sigma_b$ there, and the resulting solutions are symmetric about the midpoint, whose position therefore does not move.

We find that equations \eqref{eq:1D_eqns} possess two special solutions, both obtainable in closed form: a steady state corresponding to a tissue that is not growing, and another in which the tissue grows perennially.

For the steady state, i.e. $v=0$, the displacement field is $u^\star = (\sigma_b/E)\,x$. Note that the size of a non-growing tissue is not uniquely specifiable by the steady-state solution. However, it may be obtainable via the temporal evolution of the tissues. Hence, relative changes in the length are physically meaningful. We find that the steady-state size $L^\star$ of the tissue can be expressed as
\begin{align}
\label{eq:length_imposed_stress_model}
\frac{L^\star}{L_0} &= \left(1- \frac{\sigma_b}{E} \right)^{-1},
\end{align}
where $L_0$ is the ``initial size'' of the tissue. The expression on the right-hand side of the above equation can be interpreted as a growth-factor. Evidently this growth-factor diverges as $\sigma_b \to E$, and becomes negative, i.e. unphysical, beyond it. The reason lies in the constitutive law. Measured on the current configuration, the strain in extension is bounded, and with it the elastic stress; since the steady state has $\partial_x u^\star = \sigma_b/E$, it exists only while $\sigma_b < E$. A tension exceeding $E$ cannot be balanced at any strain, and the tissue has no choice but to flow indefinitely, giving instead the perennially growing solution obtained below. No comparable bound exists under compression, which is why the model admits both bounded and unbounded growth but only bounded de-growth. The bound is a property of the constitutive law rather than of the kinematics; the strain measure, its frame-indifference, and the class of elastic responses that share this property are discussed in \cite{SI}. The growth factor also determines the linear stability of the steady state. Writing $\epsilon = 1 - \sigma_b/E$, the steady-state size diverges as $\epsilon^{-1}$ while the rate at which it is approached vanishes as $\epsilon^{3}$ \cite{SI}, so that close to the threshold the tissue grows to an ever larger size and takes ever longer to reach it.

In the perennially growing phase, when $\sigma_b>E$, we find that the displacement field is
\begin{align}
u^\ast = x,
\end{align}
where $x \in \left(-\tfrac{L^\ast(t)}{2}, \tfrac{L^\ast(t)}{2} \right)$ with $L^\ast(t)$ being the time-dependent domain length. The velocity field is given by
\begin{align}
\label{eq:v_ugp_toy_model}
v^\ast(x,t) = \frac{\sigma_b - E}{Z} \operatorname{sech} \left(\frac{L^\ast(t)}{2 \ell}\right) \sinh \left(\frac{x}{\ell}\right), 
\end{align}
where $\ell = \sqrt{\eta/\gamma}$, $Z = \sqrt{\eta\gamma}$, and $\tau = \eta/E$. The impedance $Z$ has dimensions of stress per unit velocity and converts the excess stress $\sigma_b - E$ into an edge speed. Considering the motion of the end points of the tissue, the asymptotic growth rate $(dL/dt)^\ast$ satisfies:
\begin{align}
\left(\frac{dL}{dt}\right)^\ast &= \frac{2(\sigma_b - E)}{Z} \tanh \left(\frac{L^\ast(t)}{2 \ell}\right).
\label{eq:growth_rate_toy_model}
\end{align}
Since the same stress is applied at the two boundaries, equation \eqref{eq:v_ugp_toy_model} shows that $v$ vanishes at the midpoint $x=0$. Material points there separate exponentially at first, but only transiently: the local strain rate decays as the tissue lengthens, and over the entire history of the tissue two neighbouring points at the midpoint separate by no more than a finite factor \cite{SI}. By contrast, $v^\ast$ attains its largest magnitude at the boundaries, and for $L^\ast(t) \gg \ell$ the strain rate $\partial_x v^\ast$ is confined to within a distance $\sim\ell$ of them: the tissue interior is asymptotically quiescent and growth occurs at the edges. The growth rate then approaches the constant $2(\sigma_b - E)/Z$, so that the domain grows linearly in time despite the spatially nonuniform velocity.

The steady state is exact, and the second solution exact on the growing branch $u^\ast = x$, but the transient that connects a given initial condition to one of them is not, and neither is the behaviour at the transition itself. We therefore performed explicit numerical integrations of \eqref{eq:1D_eqns}, which show that the tissue approaches one or other of the two special solutions according to the value of $\sigma_b/E$ \cite{SI}.

The steady state is thus reminiscent of determinate growth and the perennially growing state of indeterminate growth, both in the shape of the growth curve and in the response to conditions, since planaria grow or shrink according to the availability of nutrients \cite{RinkDev2019, baguna1981}. That the two appear as distinct phases of one continuum model, separated by a balance between the applied stress $\sigma_b$ and the material parameter $E$, is the central point of this section.

\subsection{Proportionate growth} We now extend the above model to consider an inhomogeneous tissue composed of two parts with different material parameters $E_\alpha$, $\gamma_\alpha$ and $\eta_\alpha$, where $\alpha=\{l,r\}$. We denote the position of the interface between these parts by $x_I(t)$, and the edges of the tissue by $x_l(t)$ and $x_r(t)$. The tissue size is thus $L = x_r - x_l = L_l + L_r$ where $L_l = x_I - x_l$ and $L_r = x_r - x_I$ are the lengths of the two parts. Both parts of the domain obey equations \eqref{eq:1D_eqns}, with $\sigma = \sigma_b$ at both edges $x_l$ and $x_r$ as before. Additionally, since the interface is a material surface, $u$, $v$, and $\sigma$ are continuous across it.

We find three special solutions for the inhomogeneous tissue: (i) both parts of the tissue reach a non-growing steady state, (ii) both parts grow perennially, and (iii) one part reaches a non-growing steady-state while the other grows indefinitely.

The steady state solution, i.e., $v^\star_\alpha = 0$, exists when $\sigma_b < \min(E_l, E_r)$. In this case the displacement field is $u^\star = (\sigma_b/E_\alpha)\,x$ for $x$ in part $\alpha$, where, since only relative positions carry physical meaning, we have placed the origin at the interface, so that $x_I = 0$ and $x_l<0<x_r$. Each part therefore stretches by its own factor, $L^\star_\alpha = (1-\sigma_b/E_\alpha)^{-1} L_{\alpha 0}$, and the steady-state length is
\begin{align}
\label{eq:length_heterogeneous_toy_model}
L^\star = \left(1-\frac{\sigma_b}{E_l}\right)^{-1} L_{l0} + \left(1-\frac{\sigma_b}{E_r}\right)^{-1} L_{r0},
\end{align}
where $L_{\alpha 0}$ are the initial lengths of the two parts.

The second case, where both parts grow perennially, requires $\sigma_b>\max(E_l, E_r)$. Here $u^\ast = x$ across the entire tissue, as for the homogeneous one, and the edge velocities $v_\alpha \equiv \dot{x}_\alpha$ and interface velocity $v_I \equiv \dot{x}_I$ follow in closed form from force balance in each part together with continuity of $v$ and $\sigma$ at the interface; the growth rate $\dot{L}^\ast = v_r - v_l$ reduces to \eqref{eq:growth_rate_toy_model} when the two parts are identical \cite{SI}.

In the limit that each part is much longer than its own hydrodynamic length, $L_\alpha \gg \ell_\alpha = \sqrt{\eta_\alpha/\gamma_\alpha}$, the three velocities reduce to $\dot x_l = -w_l$, $\dot x_r = w_r$ and $\dot x_I = \Delta$, where
\begin{align}
\label{eq:edge_speeds}
w_\alpha \equiv \frac{\sigma_b - E_\alpha}{Z_\alpha},
\qquad
\Delta \equiv \frac{E_r-E_l}{Z_l+Z_r},
\end{align}
and $Z_\alpha = \sqrt{\eta_\alpha \gamma_\alpha}$ is the impedance of part $\alpha$.
Each edge thus advances at exactly the speed $w_\alpha$ that a homogeneous tissue with the material parameters of that part would have, while the interface drifts at a speed $\Delta$ set by the mismatch in elastic moduli. The two parts consequently grow at rates $\dot L_l = w_l + \Delta$ and $\dot L_r = w_r - \Delta$, so that $\dot L^\ast = w_l + w_r$. Since both parts grow linearly, their length fractions approach the ratios of these rates, $L_\alpha/L \to \dot L_\alpha/\dot L^\ast$, that is
\begin{align}
\label{eq:fractional_length_heterogeneous_toy_model}
\frac{L_l}{L} = \frac{Z_r}{Z_l+Z_r} + \frac{2\Delta}{\dot L^\ast},
\quad
\frac{L_r}{L} = \frac{Z_l}{Z_l+Z_r} - \frac{2\Delta}{\dot L^\ast}.
\end{align}
A composite tissue with $\sigma_b>\max(E_l, E_r)$ thus grows with its length fractions frozen in time, set by the ratio of impedances and shifted by $2\Delta/\dot L^\ast$ when the moduli differ. For both fractions to remain positive the interface must not outrun either edge, which requires $-w_l < \Delta < w_r$; when this is violated the softer part progressively consumes the stiffer one and the proportions are not maintained.

We emphasise that frozen fractions are not the same thing as preserved proportions. The values in \eqref{eq:fractional_length_heterogeneous_toy_model} are fixed by the material parameters alone and bear no relation to the proportions $L_{\alpha 0}/L_0$ the tissue started with: a composite tissue generically drifts away from its initial proportions and then holds whatever proportions the parameters dictate. Both notions should in turn be distinguished from the scaling of a chemical pattern \cite{AguilarHidalgo2018, Werner2015}, in which it is the decay length of a morphogen profile that keeps a fixed ratio to the tissue length. That mechanism preserves positional information, and so allows cell identities to be re-specified as the tissue grows; what is at stake here is instead the length occupied by parts whose identity is already fixed. The two address different problems and neither subsumes the other. Proportionate growth in the biological sense is the stronger requirement that the initial proportions be the ones maintained. Writing $\varphi = L_{l0}/L_0$, this amounts to the single condition $\Delta = \varphi \, w_r - (1-\varphi) \, w_l$ on the six material parameters, so that a five-parameter family of distinct tissues preserves any given set of proportions. It is this redundancy that can be exploited, and the most transparent way to do so is to match impedances. With equal elastic moduli the interface does not drift, $\Delta = 0$, and the requirement collapses to $Z_l/Z_r = L_{r0}/L_{l0}$: the impedances must be in inverse proportion to the initial lengths. Since an edge advances at $w_\alpha = (\sigma_b - E)/Z_\alpha$, the more heavily damped part elongates more slowly, and matching it to the initially shorter part keeps the two in step, so that the tissue grows while preserving the proportions it began with \cite{SI}. Figure~\ref{fig:prop_growth} shows the generic, unmatched case, in which the fractions freeze at values set by the material parameters irrespective of where they started.

Finally, when $\min(E_l, E_r)<\sigma_b<\max(E_l, E_r)$, the part of the tissue for which $E_\alpha>\sigma_b$ stops growing, while the other grows indefinitely.

\begin{figure}[t]
\includegraphics[width=\linewidth]{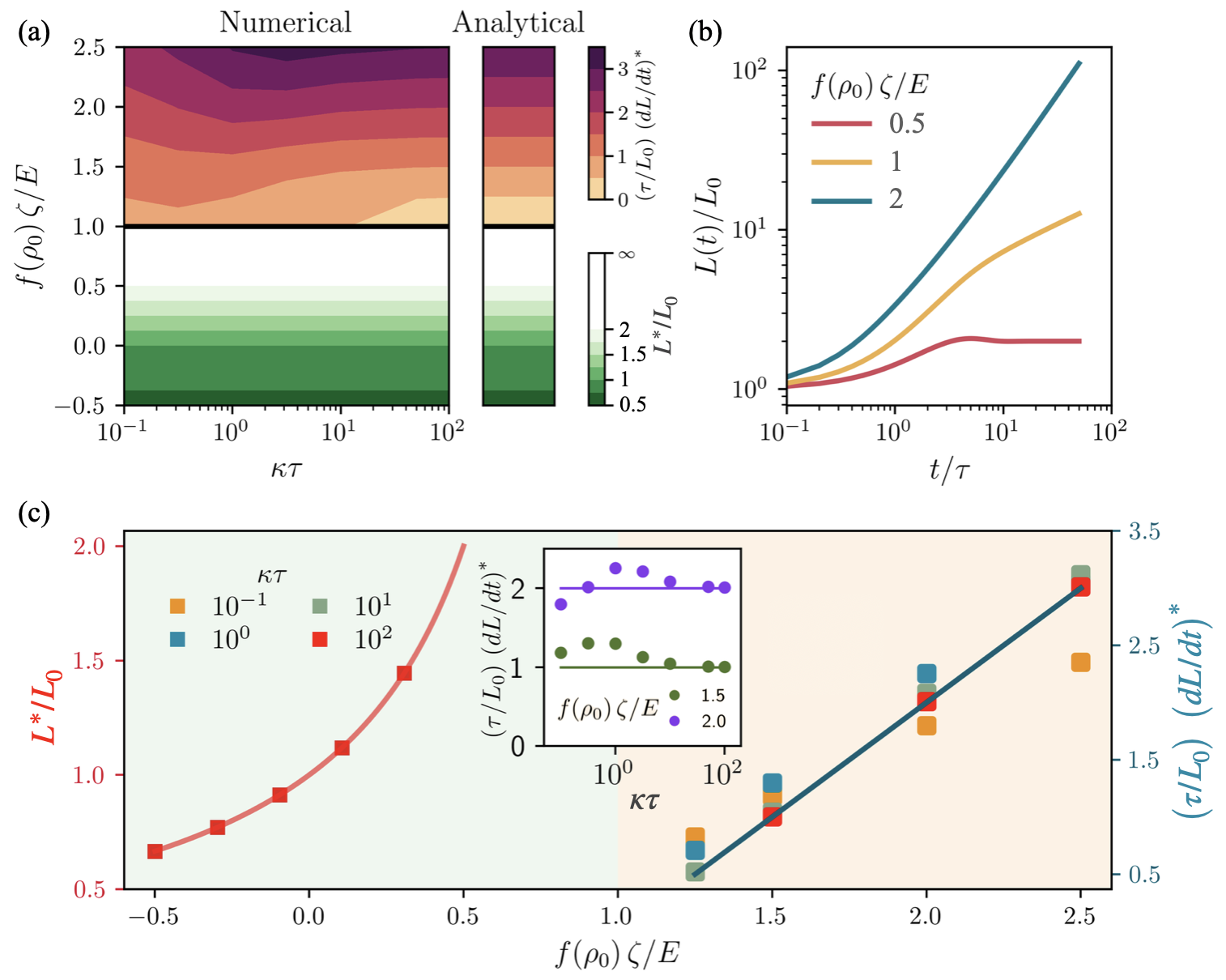}
\caption{Autonomous growth driven by activity. (a) Phase diagram in the $\zeta f(\rho_0)/E$--$\kappa\tau$ plane, numerics beside analytics. Below the transition the contours (green) are of the steady-state length $L^\star/L_0$; above it, of the asymptotic growth rate $(\tau/L_0)(dL/dt)^\ast$. (b) Tissue length against time for three values of $\zeta f(\rho_0)/E$. (c) Steady-state length (left axis) and asymptotic growth rate (right axis) against $\zeta f(\rho_0)/E$; curves are analytical, symbols numerical, for several $\kappa\tau$. The steady-state length is independent of $\kappa\tau$; the growth rate is not. Inset: the growth rate approaches its analytical value (horizontal lines) as $\kappa\tau$ grows.}
\label{fig:model_with_activity}
\end{figure}

\section{Autonomous growth driven by activity} For a tissue evolving autonomously, $\sigma_b$ should itself be determined self-consistently by other dynamical variables: the cell density, signalling morphogens, or the anisotropic degrees of freedom. We take the first of these, and ask whether the two special solutions persist when the driving is generated internally, $\tens{\sigma}_\mathrm{active} \neq 0$, rather than imposed at the boundary. They do.

The cell number density field $\rho$ and the locations where it is non-zero define the extent of the tissue. It evolves according to
\begin{align}
\label{eq:density_evolution}
    D_t \rho = - (\nabla \cdot \vect v) \, \rho + \kappa \rho \left(1-\frac{\rho}{\rho_0}\right),
\end{align}
where $\kappa$ is the cellular birth--death rate and $\rho_0$ is the carrying capacity for the cell density. We have chosen logistic reaction to mimic cell division and apoptosis. Spatial variations of the cell density $\rho$ within the tissue arising from cell birth and death events generates local pushing and pulling forces. This effect can be encapsulated in an isotropic active stress $\tens{\sigma}_{\mathrm{active}} = -\zeta \, f(\rho) \, \tens{I}$, where $\zeta$ is the active-stress strength, and $f$ is a non-dimensional active-stress regulation function that depends on cell number density $\rho$.
To make the system fully autonomous the tissue is isolated, so that the traction vanishes on its boundary. In one dimension the governing equations are therefore \eqref{eq:material_point_evolution}, \eqref{eq:displacement_evolution} and \eqref{eq:force_balance} as before, now with total stress $\sigma = E \partial_x u + \eta \partial_x v - \zeta f(\rho)$ and $\sigma\vert_{x=\pm L(t)/2} = 0$, together with \eqref{eq:density_evolution}.

As in the model without activity, we find a spatially homogeneous and isotropic steady-state solution for a $d$-dimensional tissue of size $R^\star$ 
\begin{align}
\label{eqn:length_model_with_activity_d_dim}
    \frac{R^\star}{R_0} = \left(1 - \frac{\zeta f(\rho_0)}{d\,E}\right)^{-1}
\end{align}
where $R_0$ is the initial size of the tissue, with associated fields $\vect{u}^\star = \left[\zeta f(\rho_0)/d\,E\right]\vect{x}$ and $\rho^\star = \rho_0$. Notice the remarkable similarity of \eqref{eqn:length_model_with_activity_d_dim} and these fields to \eqref{eq:length_imposed_stress_model} and the corresponding steady state of the model with an imposed boundary stress.

There exists a perennially growing isotropic phase in $d$-dimensions, just as in the case without activity. For growth along a single axis the displacement is again $u^\ast = x$, and the velocity field is
\begin{align}
v^\ast &= \frac{\zeta f(\rho_0) - E}{Z} \operatorname{sech} \left(\frac{L^\ast(t)}{2 \ell}\right) \sinh \left(\frac{x}{\ell}\right)
\nonumber
\\
&\quad\quad -\zeta \int_{-\tfrac{L^\ast(t)}{2}}^{\tfrac{L^\ast(t)}{2}} \frac{\partial f}{\partial y} \, G(x, y)\, dy ,
\label{eq:v_activity}
\end{align}
where $G$ is the Green's function of the operator $\gamma - \eta\,\partial_x^2$ with no-flux boundary conditions \cite{SI}.
The growth rate of the domain in the perennially growing phase can be computed by evaluating $v^\ast$ at the end points and taking the limit $L^\ast \to \infty$. This leads to 
\begin{align}
\label{eq:growth_rate_1D_activity}
 \left(\frac{dL}{dt}\right)^\ast &= \frac{2\left(\zeta f(\rho_0) - E\right)}{Z} + \mathcal{O} \left(\max_{t, x} \, \partial_x \rho \right).
\end{align}
Here, we used the fact $\max_\rho |f^\prime(\rho)|$ is finite, and that $G(x, y)$ is positive and evaluates to a constant when integrated over the domain. Note again that the first part of the growth rate in this fully autonomous model resembles the growth rate of the model with no activity and an imposed stress at the boundaries. There is also an additional contribution from the spatial gradients in the density field. 

The steady state above is exact for any birth--death rate: with $\vect{v}=0$ the density equation is solved identically by $\rho = \rho_0$, whatever the value of $\kappa$. In the perennially growing phase this is no longer so, since the growth rate carries the density-gradient correction in \eqref{eq:growth_rate_1D_activity}, which is controlled only when $\kappa\tau \gg 1$ holds the density close to its setpoint. To reach small and intermediate turnover, where no such control is available, we integrated these equations numerically, by a finite-element method on a mesh deformed with the material; the choice of $f(\rho)$ and of the boundary condition on $\rho$ affects the results only quantitatively \cite{SI}. They are summarised in Fig.~\ref{fig:model_with_activity}. The transition at $\zeta f(\rho_0)/E = 1$ is reproduced, and the numerical contours approach the analytical ones as $\kappa\tau$ is increased: at large birth--death rate the density is held close to $\rho_0$, so the $\nabla\rho$ correction in \eqref{eq:growth_rate_1D_activity} is suppressed. The steady-state length is independent of $\kappa\tau$, whereas the growth rate in the unbounded phase depends on it non-monotonically.

A heterogeneous tissue with active stresses in the bulk and no applied stress at its boundary behaves exactly as the composite tissue of the previous section, with $\sigma_b \to \zeta f(\rho_0)$: the same three regimes are separated by $\min(E_l, E_r)$ and $\max(E_l, E_r)$, the steady-state lengths are given by \eqref{eq:length_heterogeneous_toy_model}, and, when spatial gradients in the density are suppressed, the long-time length fractions by \eqref{eq:fractional_length_heterogeneous_toy_model}.

\begin{figure}[t]
\centering
\includegraphics[width=\linewidth]{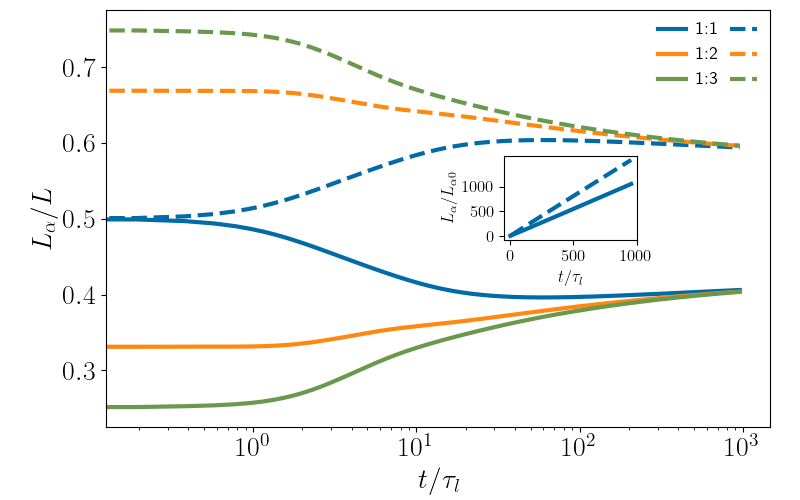}
\caption{Length fractions $L_\alpha(t)/L(t)$ of the two parts of a heterogeneous tissue in the perennially growing phase, against time in units of $\tau_l = \eta_l/E_l$ on a logarithmic scale. Solid and dashed curves are the two parts; the three colours are initial length ratios $L_{l0}:L_{r0}$ of $1\!:\!1$, $1\!:\!2$ and $1\!:\!3$, all else equal. All three initial conditions drift towards the same pair of asymptotic fractions, which are therefore set by the material parameters and not inherited. Inset: for the $1\!:\!1$ case, each part grows linearly in time once the fractions have settled. Under the impedance-matched condition $Z_l/Z_r = L_{r0}/L_{l0}$ the curves would instead stay flat at their initial values.}
\label{fig:prop_growth}
\end{figure}

\section{Discussion} We have modelled a growing tissue as an active continuum on a deforming domain, and found two special solutions -- a non-growing steady state and a perennially growing one -- selected by a single dimensionless combination, $\sigma_b/E$ under an applied boundary stress and $\zeta f(\rho_0)/E$ when driven autonomously by cell division and death. The transition between them rests on a bound on the elastic stress; no such bound exists in compression, which is why the model admits bounded and unbounded growth but only bounded de-growth. On approach to the threshold the steady-state size diverges as $(1-\sigma_b/E)^{-1}$, while the rate at which it is reached vanishes as $(1-\sigma_b/E)^{3}$. At low birth--death rate the autonomous model also oscillates as a whole, a consequence of the contractile--extensile form of the active stress (Movie~S3 \cite{SI}).

Growth is localised at the edges. Although division and death occur throughout the bulk, the velocity vanishes at the centre, and for $L \gg \ell$ the strain rate is confined to within a hydrodynamic length of the boundary, so that the length dynamics is governed by a thin peripheral layer. This is why the growth rate of a composite tissue is simply the sum of two independent edge contributions.

For an inhomogeneous tissue the length fractions freeze of their own accord, but at values set by the material parameters rather than inherited, so the proportions generically drift before settling. Preserving the initial ones is the stronger requirement, and amounts to a single scalar condition on the six material parameters, whose most transparent solution is to match impedances \cite{SI}.

Several extensions are immediate. The material parameters need not be piecewise constant, nor need they be fixed. Morphogen profiles pattern tissue mechanics \cite{Boulan2021dev, Shraiman2005-}, and a signalling field on the growing domain can be coupled in by letting $E$, $\eta$, $\gamma$ and $\zeta$ depend on it, whereupon the analysis above fixes the growth phase locally. A growth rate that follows the morphogen \emph{dynamics}, through $\partial_t \ln C$, is not the same as one that follows the morphogen \emph{level}, which is what such a coupling would give, the level fixing which side of the threshold the tissue sits on. The two are distinguishable, since a step change in morphogen would produce a transient burst of growth in the first case and a switch of phase in the second. A nematic order parameter describing oriented divisions can be added in the same way; it contributes a deviatoric active stress, which is precisely what vanishes under isotropic dilation, and is what would allow growth rates to differ between directions.

We have also held the active stress independent of the mechanical state. A homeostatic closure, in which the rate of division and death responds to the local pressure \cite{Basan2009, Ranft2010}, would determine the setpoint $\rho_0$ rather than impose it, and would give the steady state a size rather than a growth factor: the natural route to the compensatory behaviour that characterises determinate growth, and one the present model, whose uniform dilation mode is neutrally stable \cite{SI}, does not take.

Conservation laws in growth problems are agreed upon; the kinematics and constitutive laws are not \cite{ambrosi_perspectives_2011}, the contentious point being the reference configuration used to define elastic strains. One may let the reference evolve, decomposing the deformation gradient multiplicatively \cite{Goriely2007} at the cost of prescribing growth laws by hand; or endow the current body with a residual stress or a new metric \cite{Joanny2022, claussen2026elasticityreferencestatecontinuum}, or dispense with a reference altogether \cite{chen_mechanical_2024}, in either case making the constitutive relations implicit \cite{rajagopal2003implicit}. Our fixed choice is justified when the grown state is dense and when the elastic stress relaxes slowly compared with the time over which the tissue grows. It is implicit in the same sense, since the strain is measured on a domain that is itself part of the solution; in particular the material is not Kelvin--Voigt at finite strain, the spring and dashpot responding to different members of the Seth--Hill family \cite{SI}. Nor need that choice be permanent: in the perennially growing state the influence of the initial configuration vanishes asymptotically, so that it may be reset to the current one, leaving the tissue free of residual stress and able to begin anew. Such a reset is itself a relaxation mechanism, and a version of the model in which it occurs dynamically -- so that the constitutive relation is \textit{nominally} fluid at long times -- is left for future work.

Growth in more than one direction at once is a harder problem than the isotropic dilation worked out in \cite{SI}, and one we have not attempted. Once the tissue may change shape the steady state is no longer a single number, many shapes satisfy the traction-free condition, and the rates along different directions need not agree, so that the aspect ratio itself evolves. Proportionate growth then becomes a statement about shape -- whether an organ remains similar to itself -- rather than about a ratio of lengths, and we regard this as the natural next question.

\acknowledgments
We acknowledge support of the Department of Atomic Energy, Government of India, under project no. RTI4019. We used Claude (Anthropic) to assist with copyediting and drafting. All AI-assisted content was critically reviewed, verified, and edited by the authors, who take full responsibility for the accuracy, integrity, and originality of the work. We thank Mandar Inamdar, Matthias Merkel and Sundar Naganathan for fruitful discussions.

\bibliography{references}

\clearpage
\onecolumngrid

\setcounter{section}{0}
\setcounter{equation}{0}
\setcounter{figure}{0}
\renewcommand{\thesection}{S\arabic{section}}
\renewcommand{\theequation}{S\arabic{equation}}
\renewcommand{\thefigure}{S\arabic{figure}}

\begin{center}
{\large Supplementary Material for}\\[0.4em]
\textbf{\Large Growth phases of an active tissue: determinate, indeterminate,
and proportionate}\\[1em]
Jigyasa Watwani \qquad K. Vijay Kumar \qquad Vishal Vasan
\end{center}

\vspace{1em}

\noindent
This document collects the calculations underlying the results quoted in the
main text. Equation numbers prefixed by S refer to this document; unprefixed
numbers refer to the main text. Notation follows the main text throughout.

\addtocontents{toc}{\protect\setcounter{tocdepth}{2}}
\tableofcontents

\section{Invertibility of the Lagrangian map}
\label{si:kinematics}

Material points of the tissue move according to $\dot{\vect{x}} = \vect{v}(t,\vect{x})$,
with $\vect{x}(0) = \vect{X}$, and the displacement field is
$\vect{u} = \vect{x} - \vect{X}$. Two descriptions are available: a Lagrangian
one, in which fields are functions of the reference position $\vect{X}$, and an
Eulerian one, in which they are functions of the current position $\vect{x}$.
Passing between them requires the map $\vect{X} \mapsto \vect{x}(t,\vect{X})$ to
be invertible; the conditions under which it is are established below.

\subsection{The Jacobian}

Let $\tens{J} = \partial\vect{x}/\partial\vect{X}$ be the deformation gradient and
$\Lambda = \det\tens{J}$. Differentiating $\Lambda$ along a material trajectory
gives the Euler expansion formula
\begin{align}
\dot{\Lambda} = \Lambda \, (\nabla\cdot\vect{v}),
\label{si:euler_expansion}
\end{align}
whose solution is
\begin{align}
\Lambda(t) = \exp \int_0^t \left( \nabla \cdot \vect{v} \right) dt^\prime ,
\label{si:jacobian}
\end{align}
the integral being taken along the trajectory. A uniform bound on the divergence
suffices for $\tens{J}$ to be non-singular. If $|\nabla\cdot\vect{v}| \leq M$
with $M$ independent of $t$, then $e^{-Mt} \leq \Lambda(t) \leq e^{Mt}$, so that
$\Lambda$ remains finite and strictly positive at any finite time, the map
$\vect{X}\mapsto\vect{x}$ is invertible, and the Eulerian description is
available. The bound does not survive the limit $t \to \infty$: a bounded but
strictly positive $\nabla\cdot\vect{v}$ suffices to make $\Lambda$ diverge, which
is what occurs in the perennially growing phase.

\subsection{Affine displacement fields}
\label{si:affine}

The two asymptotic states quoted in the main text follow from the affine
displacement fields compatible with the kinematics. With $\vect{x}$ an Eulerian
position vector, let
\begin{align}
\vect{u} = \tens{A} \cdot \vect{x} ,
\end{align}
with $\tens{A}$ a constant tensor. Substituting into
$D_t\vect{u} = \vect{v}$ and using $D_t \vect{x} = \vect{v}$ gives
\begin{align}
\left( \tens{A} - \tens{I} \right) \cdot \vect{v} = \vect{0} ,
\label{si:affine_condition}
\end{align}
where $\tens{I}$ is the identity tensor in $d$ dimensions. Equation
\eqref{si:affine_condition} is satisfied in two ways:

\begin{itemize}
\item[(i)] $\vect{v} = \vect{0}$, with $\tens{A}$ arbitrary. The domain is
static and the displacement field is whatever the force balance and the
boundary condition select.
\item[(ii)] $\vect{v} \neq \vect{0}$ lies in the kernel of
$\tens{A}-\tens{I}$, that is, $\vect{v}$ is an eigenvector of $\tens{A}$ with
unit eigenvalue.
\end{itemize}

Case (ii) singles out a direction, which an isotropic tissue does not possess:
with no preferred direction, every direction must be an eigenvector with unit
eigenvalue, so that $\tens{A} = \tens{I}$ and $\vect{u} = \vect{x}$. The velocity
is then left undetermined by the kinematics and is fixed instead by force
balance, as computed in the main text.

The two cases correspond respectively to an asymptotic state in which $\Omega_t$
is static and to one in which it deforms perennially. Neither is an assumption
about the dynamics: together they exhaust the affine possibilities, and the
numerical solutions of the main text confirm that one or the other is selected
according to the value of the control parameter. They do not exhaust the
solutions of the full problem: at small $\kappa\tau$ the tissue instead
oscillates as a whole (Movie~S3), a state that is not of affine form.

\subsection{The degenerate configuration \texorpdfstring{$\vect{u}=\vect{x}$}{u=x}}

Since $\vect{u}(t,\vect{x}) = \vect{x} - \vect{X}$, we have
$\partial\vect{X}/\partial\vect{x} = \tens{I} - \nabla\vect{u}$, and therefore
\begin{align}
\tens{J} = \left( \tens{I} - \nabla\vect{u} \right)^{-1} .
\label{si:deformation_gradient}
\end{align}
The affine analysis of Sec.~\ref{si:affine} singles out two special
configurations: $\vect{v} = \vect{0}$, and $\vect{u} = \vect{x}$. The second is
precisely $\nabla\vect{u} = \tens{I}$, at which \eqref{si:deformation_gradient} is
singular. Equivalently $\vect{X} = \vect{0}$: every material point of the tissue
traces back to a single point of the initial configuration, and all information
about $\Omega_0$ has been lost.

In the perennially growing state $\nabla\cdot\vect{v}$ remains bounded and positive, so by \eqref{si:jacobian}
$\Lambda$ diverges only as $t\to\infty$; the configuration $\vect{u} = \vect{x}$
is approached asymptotically and is never attained at finite time. The
perennially growing state is therefore indeterminate in two ways: the size
grows without bound, and the reference configuration ceases to influence the
dynamics.

The second sense underlies the generalisation discussed in the main text, in
which the reference configuration is reset to the current one and the dynamics
restarted from a stress-free state.

\section{The elastic constitutive law}
\label{si:constitutive}

\subsection{Strain measure and objectivity}

The elastic stress of the main text is built from the symmetrised gradient
\begin{align}
\tens{\varepsilon} = \tfrac{1}{2}\left[ \nabla\vect{u} + (\nabla\vect{u})^{\mathsf{T}} \right]
= \tens{I} - \tfrac{1}{2}\left( \tens{J}^{-1} + \tens{J}^{-\mathsf{T}} \right),
\label{si:strain}
\end{align}
where the second equality follows from \eqref{si:deformation_gradient}. This is
not, strictly, a frame-indifferent measure of strain at finite deformation.
Under a rigid rotation $\vect{x} \to \tens{Q}\cdot\vect{x}$ the deformation
gradient transforms as $\tens{J} \to \tens{Q}\cdot\tens{J}$, so that
$\tens{J}^{-1} \to \tens{J}^{-1}\cdot\tens{Q}^{\mathsf{T}}$, and the combination
appearing in \eqref{si:strain} does not transform as a second-rank tensor
should. The objective $m=-2$ member of the Seth--Hill family is instead the
Almansi strain
\begin{align}
\tens{e} = \tfrac{1}{2}\left[ \tens{I} - \left( \tens{J}\tens{J}^{\mathsf{T}}\right)^{-1} \right],
\label{si:almansi}
\end{align}
which is built from the left Cauchy--Green tensor $\tens{J}\tens{J}^{\mathsf{T}}
\to \tens{Q}\tens{J}\tens{J}^{\mathsf{T}}\tens{Q}^{\mathsf{T}}$ and is therefore
objective. The two agree to linear order in $\nabla\vect{u}$.

We adopt \eqref{si:strain} rather than \eqref{si:almansi}. All explicit results
below are obtained in one spatial dimension, where rotations are absent and the
distinction does not affect the structure of the solutions. Both measures are
bounded above under extension, so the mechanism responsible for the transition
is common to them and only the numerical value of the threshold differs;
Sec.~\ref{si:which_laws} identifies the constitutive laws that share this
property.

\subsection{The bound on the elastic stress}

In one dimension let $\lambda = \partial x/\partial X > 0$ denote the stretch.
From \eqref{si:deformation_gradient},
\begin{align}
\partial_x u = 1 - \frac{1}{\lambda} ,
\label{si:strain_1d}
\end{align}
which is the $m=-1$ member of the Seth--Hill family of generalised strain
measures, $\mathcal{E}^{(m)} = (\lambda^m-1)/m$. It is bounded above,
\begin{align}
\partial_x u < 1 \qquad \mbox{for every} \quad \lambda > 0 ,
\label{si:saturation}
\end{align}
with the bound approached only as $\lambda \to \infty$. The elastic stress
therefore cannot exceed $E$, however far the tissue is stretched. An applied
tension $\sigma_b > E$ cannot be balanced elastically at any strain, and the
tissue has no alternative but to flow indefinitely. This is the origin of the
threshold at $\sigma_b = E$ in the model with an imposed boundary stress, and of
its autonomous counterpart, which is $\zeta f(\rho_0) = E$ in one dimension and
$\zeta f(\rho_0) = d\,E$ in $d$ dimensions (Sec.~\ref{si:ddim}).

\subsection{Which elastic responses admit a threshold}
\label{si:which_laws}

Within the Seth--Hill family, a linear law $\sigma = E \, \mathcal{E}^{(m)}$
admits a threshold if and only if $m<0$, since $\mathcal{E}^{(m)} \to -1/m$ as
$\lambda \to \infty$ for negative $m$ and diverges otherwise. The measure used
here, $m=-1$, gives $\sigma < E$, and the Almansi measure $m=-2$, of
one-dimensional form $\tfrac{1}{2}(1-\lambda^{-2})$, gives $\sigma < E/2$. The
logarithmic ($m=0$), Biot ($m=1$) and Green ($m=2$) measures give $\ln\lambda$,
$\lambda-1$ and $\tfrac{1}{2}(\lambda^2-1)$ respectively, all unbounded; under
these no applied tension is too large to be balanced, and the tissue is
determinate at every $\sigma_b$.

The transition is therefore a consequence of the constitutive law rather than of
the kinematics. What it requires is that the Cauchy stress the material can carry
in extension be bounded, the particular measure fixing only the numerical value
of the threshold. The requirement is a statement about the tissue: a material
whose elastic stress ceases to grow beyond some extension stops storing energy
and begins to flow, as tissues do when cells rearrange through T1 transitions
rather than stiffening indefinitely. Thus $E$ is the stress above
which the tissue can no longer respond elastically, and the
determinate--indeterminate transition is the statement that a tissue holds a
fixed size only while its activity remains below it. The $m=-1$ law is the
simplest closure with this property.

In compression no such bound exists, for any $m$: as $\lambda \to 0$,
\eqref{si:strain_1d} gives $\partial_x u \to -\infty$, so a compression of any
magnitude can be balanced by a sufficiently large contraction. This asymmetry
between extension and compression is why the model admits both bounded and
unbounded growth but only bounded de-growth.

\subsection{The spring and the dashpot refer to different measures}

The elastic and viscous stresses of the main text are referred to different
strain measures, so that the material is not a Kelvin--Voigt solid at finite
strain. Writing $\lambda = \partial x/\partial X$,
\begin{align}
\dot{\lambda} = \frac{\partial}{\partial X}\left( \frac{\partial x}{\partial t} \bigg\vert_X \right)
= \frac{\partial v}{\partial X} = \lambda \, \frac{\partial v}{\partial x} ,
\end{align}
so that $\partial_x v = \dot{\lambda}/\lambda = d\mathcal{E}^{(0)}/dt$ with
$\mathcal{E}^{(0)} = \ln\lambda$. The dashpot therefore responds to the rate of
the logarithmic ($m=0$) strain while the spring responds to the $m=-1$ strain.
The two elements therefore cannot be assembled into a standard rheological model
at finite deformation, and the elastic stress is not derivable, as the
work-conjugate of $\mathcal{E}^{(-1)}$, from a strain energy. The constitutive
relation is a phenomenological closure chosen so that the extensional response
is bounded, and is not presented as the linearisation of a hyperelastic solid.

\section{Linear stability of the homogeneous steady state}
\label{si:stability}

We map the moving domain to a fixed one by setting
$s = \left[ x + L(t)/2 \right]/L(t)$, so that $s \in [0,1]$ at all times. Time
is not rescaled; we write $\partial_t\vert_s$ for the time derivative taken at
fixed $s$, which is not the same as the derivative at fixed $x$. The
displacement and velocity equations become
\begin{align}
\left. \frac{\partial u}{\partial t} \right\vert_s
+ \frac{\partial u}{\partial s}\frac{1}{L}\left( v - \dot{L}\left(s-\frac{1}{2}\right)\right) &= v ,
\\
\gamma v &= \frac{E}{L^2}\frac{\partial^2 u}{\partial s^2}
+ \frac{\eta}{L^2}\frac{\partial^2 v}{\partial s^2} ,
\end{align}
with boundary condition
\begin{align}
\left. \frac{E}{L}\frac{\partial u}{\partial s}
+ \frac{\eta}{L}\frac{\partial v}{\partial s} \right\vert_{\{0,1\}} = \sigma_b .
\end{align}
The combination appearing in the advective term is just the velocity of the
rescaled coordinate,
\begin{align}
\frac{1}{L}\left( v - \left(s-\frac{1}{2}\right)\dot{L} \right) = \frac{ds}{dt} ,
\end{align}
and since the end points are stationary in the new variable,
\begin{align}
\frac{\dot{L}}{2} = v(1,t) = -v(0,t) .
\end{align}

We now perturb about the steady state $v^\star = 0$,
$u^\star = \sigma_b L^\star s/E$, $L^\star = L_0 (1-\sigma_b/E)^{-1}$, retain
terms to linear order, shift the displacement perturbation so that it obeys
homogeneous boundary conditions, and expand the result in a cosine basis,
\begin{align}
\widetilde{\delta u} = \frac{\delta u_0}{2} + \sum_{n=1}^N \delta u_n \cos(n\pi s),
\qquad
\delta v = \frac{\delta v_0}{2} + \sum_{n=1}^N \delta v_n \cos(n\pi s) .
\end{align}
Eliminating $\delta v_n$ in favour of $\delta u_n$ gives
\begin{align}
\dot{\delta u}_m = \left( \frac{\sigma_b}{E} - 1 \right) \alpha_m^2 \, \delta u_m ,
\qquad
\dot{\delta u}_0 = 0 ,
\label{si:stability_modes}
\end{align}
with
\begin{align}
\alpha_m^2 = \frac{E\,m^2\pi^2}{\gamma {L^\star}^2 + \eta \, m^2\pi^2} > 0 .
\end{align}

The mode $m=0$ is neutrally stable. This is the statement, made in the main
text, that the steady state does not fix the size of the tissue: a uniform
dilation carries one steady state into another. All remaining modes decay when
$\sigma_b < E$ and grow when $\sigma_b > E$. The ratio $\sigma_b/E$ is therefore
the control parameter for the stability of the homogeneous steady state, and the
approach to $L^\star$ in the determinate phase is exponential, at a rate set by
the slowest decaying mode, $\left| \sigma_b/E - 1 \right| \alpha_1^2$.

That rate vanishes faster than linearly as the threshold is approached, because
$\alpha_1^2$ depends on $L^\star$, which itself diverges. Writing
$\epsilon = 1 - \sigma_b/E$, so that $L^\star = L_0/\epsilon$, the friction term
dominates the denominator of $\alpha_1^2$ once
$\gamma L_0^2/\epsilon^2 \gg \eta\pi^2$, and
\begin{align}
\alpha_1^2 \to \frac{E\pi^2}{\gamma L_0^2} \, \epsilon^2 ,
\qquad
\left| \frac{\sigma_b}{E} - 1 \right| \alpha_1^2
\to \frac{E\pi^2}{\gamma L_0^2} \, \epsilon^3 .
\label{si:rate_scaling}
\end{align}
The steady-state size therefore diverges as $\epsilon^{-1}$ while the rate at
which it is reached vanishes as $\epsilon^{3}$: the two do not share an
exponent, the difference arising entirely from the $L^\star$-dependence of
$\alpha_1^2$.

\section{Separation of material points at the midpoint}
\label{si:midpoint}

In the perennially growing phase the velocity field of a homogeneous tissue is
\begin{align}
v^\ast(x,t) = \frac{\sigma_b - E}{Z} \operatorname{sech}
\left( \frac{L^\ast(t)}{2\ell} \right) \sinh \left( \frac{x}{\ell} \right) ,
\label{si:v_growing}
\end{align}
which vanishes at the midpoint $x=0$, since the same stress is applied at the
two boundaries. Expanding about that point, $\sinh(x/\ell) \simeq x/\ell$, gives
a locally linear velocity $v^\ast \simeq k(t)\,x$ with strain rate
\begin{align}
k(t) = \frac{\sigma_b - E}{Z\ell} \operatorname{sech}
\left( \frac{L^\ast(t)}{2\ell} \right)
= \frac{\sigma_b-E}{\eta} \operatorname{sech}\left( \frac{L^\ast(t)}{2\ell} \right) ,
\label{si:midpoint_rate}
\end{align}
where we used $Z\ell = \sqrt{\eta\gamma}\sqrt{\eta/\gamma} = \eta$. Two material
points straddling the midpoint therefore separate exponentially, at the
instantaneous rate $k$.

The separation is nonetheless bounded, because $k$ decays as the tissue
lengthens. The total logarithmic separation accumulated over the entire history
of the tissue is $\int_0^\infty k \, dt$, which may be evaluated by changing
variables from $t$ to $L^\ast$ using the growth law
$(dL/dt)^\ast = 2(\sigma_b-E) Z^{-1} \tanh(L^\ast/2\ell)$. With
$Z/\eta = \ell^{-1}$ and $\operatorname{sech}/\tanh = 1/\sinh$,
\begin{align}
k \, dt = \frac{1}{2\ell} \operatorname{cosech}
\left( \frac{L^\ast}{2\ell} \right) dL^\ast
= \operatorname{cosech}(s) \, ds ,
\qquad s = \frac{L^\ast}{2\ell} ,
\end{align}
and since $\int \operatorname{cosech} s \, ds = \ln \tanh (s/2)$, integrating
from the initial length $L_0$ to $L^\ast \to \infty$ gives
\begin{align}
\int_0^\infty k \, dt = - \ln \tanh \left( \frac{L_0}{4\ell} \right) .
\label{si:midpoint_integral}
\end{align}
Two material points near the midpoint thus separate by the finite factor
$\coth(L_0/4\ell)$ over the whole life of the tissue, however long it grows.
From \eqref{si:midpoint_rate} the exponential regime lasts only while
$L^\ast \lesssim \ell$; thereafter $\operatorname{sech}(L^\ast/2\ell)$ is
exponentially small and the interior is quiescent, with the strain rate confined
to within $\sim\ell$ of the two edges.

\section{Time-dependent solutions under an applied stress}
\label{si:1d_numerics}

\begin{figure}[htbp]
\centering
\includegraphics[width=0.75\linewidth]{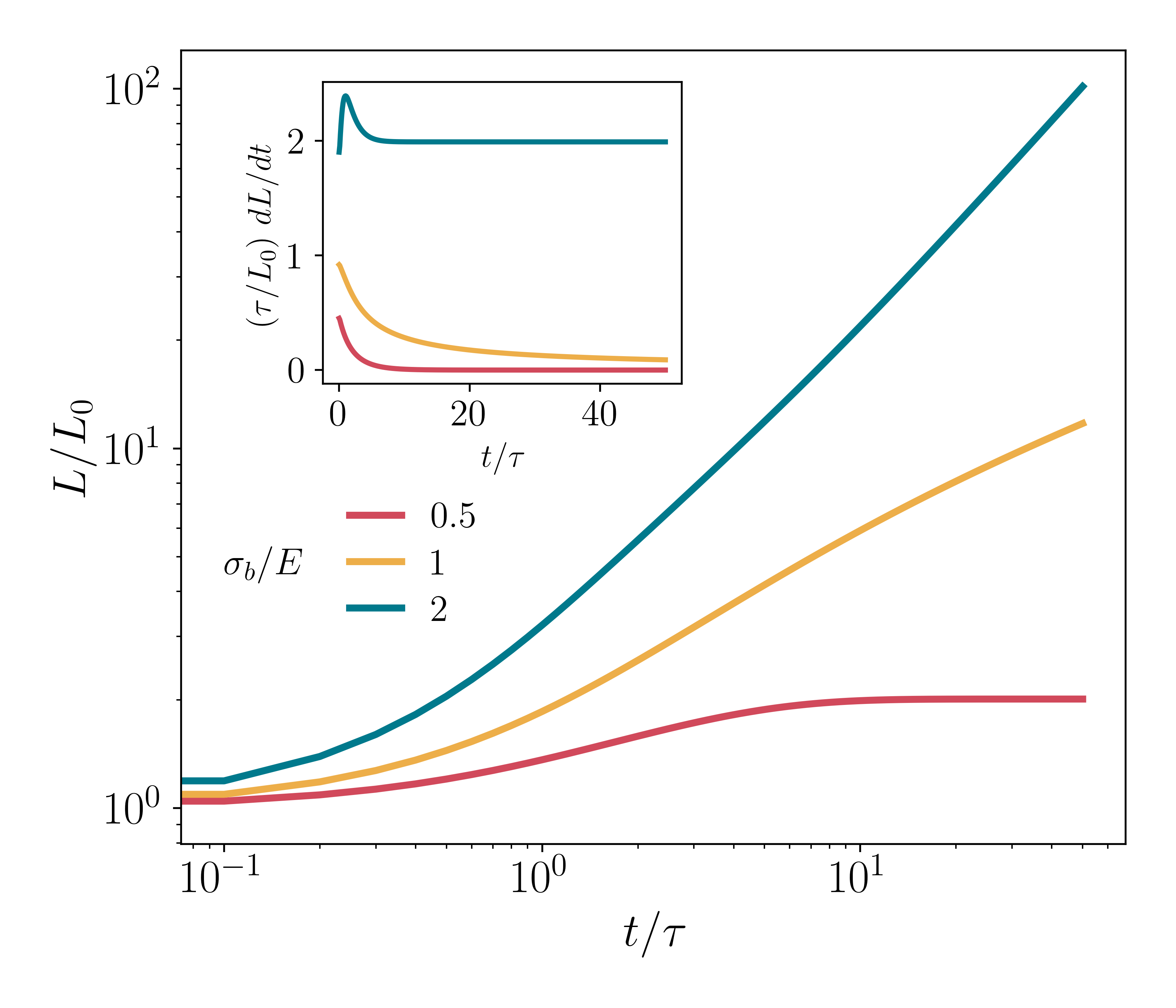}
\caption{Growth of a homogeneous tissue under an applied boundary stress,
from numerical integration of the one-dimensional equations, for three values of
$\sigma_b/E$. Below the transition the length saturates at the steady-state
value; above it the tissue grows without bound and asymptotically linearly.
Inset: the growth rate $(\tau/L_0)\,dL/dt$, decaying to zero in the bounded
phase and approaching $2(\sigma_b-E)/Z$ in the unbounded one. At
$\sigma_b/E=1$ the length still diverges, but too slowly to reach a linear
regime.}
\label{si:fig:1D_lengths_velocities}
\end{figure}

The two solutions of the imposed-stress model quoted in the main text are
asymptotic: they describe the states the tissue settles into, not the transient
by which it gets there, and neither applies at the transition itself. Figure
\ref{si:fig:1D_lengths_velocities} shows time-dependent solutions obtained by
integrating the governing equations numerically, by the method of
Sec.~\ref{si:numerics}.

Below the transition the length saturates at the value predicted analytically, and the approach is exponential, at the rate given
by the slowest decaying mode of Sec.~\ref{si:stability}. Above it the growth rate
settles onto the predicted constant $2(\sigma_b-E)/Z$. At the threshold
$\sigma_b = E$ neither special solution applies: the length still grows without
bound, but the growth rate decays slowly and no linear regime is reached within
the time simulated.

\section{Velocity field of the two-part tissue}
\label{si:twopart}

Consider the inhomogeneous tissue of the main text: a left part occupying
$[x_l, x_I]$ with parameters $E_l, \eta_l, \gamma_l$ and a right part occupying
$[x_I, x_r]$ with parameters $E_r, \eta_r, \gamma_r$. In the perennially growing
phase $u^\ast = x$ throughout, so $\partial_x u = 1$ and the stress in part
$\alpha$ is $\sigma_\alpha = E_\alpha + \eta_\alpha \partial_x v$. Force balance
$\partial_x \sigma = \gamma_\alpha v$ then reduces to
\begin{align}
\ell_\alpha^2 \, \partial_x^2 v = v ,
\qquad
\ell_\alpha = \sqrt{\eta_\alpha/\gamma_\alpha} ,
\end{align}
in each part separately. Writing the solution in the left part as
$v = A\cosh\theta + B\sinh\theta$ with $\theta = (x-x_l)/\ell_l$, and in the
right part as $v = v_I \cosh\theta^\prime + D\sinh\theta^\prime$ with
$\theta^\prime = (x-x_I)/\ell_r$, the stresses read
\begin{align}
\sigma_l = E_l + Z_l \left( A\sinh\theta + B\cosh\theta \right),
\qquad
\sigma_r = E_r + Z_r \left( v_I \sinh\theta^\prime + D\cosh\theta^\prime \right),
\end{align}
where $Z_\alpha = \eta_\alpha/\ell_\alpha = \sqrt{\eta_\alpha\gamma_\alpha}$.

The four constants are fixed by $\sigma = \sigma_b$ at $x_l$ and at $x_r$,
together with continuity of $v$ and of $\sigma$ at the interface. Solving and
writing $S_\alpha = \operatorname{sech}(L_\alpha/\ell_\alpha)$,
$T_\alpha = \tanh(L_\alpha/\ell_\alpha)$ and
$\mathcal{D}^{-1} = Z_l T_l + Z_r T_r$, the edge velocities
$v_\alpha \equiv \dot{x}_\alpha$ and the interface velocity
$v_I \equiv \dot{x}_I$ are
\begin{align}
\begin{bmatrix} v_l \\ v_I \\ v_r \end{bmatrix}
= \mathcal{D}
\begin{bmatrix}
1 + \tfrac{Z_r}{Z_l}T_l T_r & S_l & S_l S_r \\
S_l & 1 & S_r \\
S_l S_r & S_r & 1 + \tfrac{Z_l}{Z_r}T_l T_r
\end{bmatrix}
\begin{bmatrix} E_l - \sigma_b \\ E_r - E_l \\ \sigma_b - E_r \end{bmatrix} .
\label{si:matrix}
\end{align}
The total length $L = L_l + L_r$ therefore grows at the rate
$\dot{L}^\ast = v_r - v_l$, that is
\begin{align}
\frac{dL^\ast}{dt} = \mathcal{D} \Big[
&\left( (1-S_l)(1+S_r) + \tfrac{Z_r}{Z_l}T_l T_r \right)(\sigma_b - E_l)
\nonumber \\
+ &\left( (1+S_l)(1-S_r) + \tfrac{Z_l}{Z_r}T_l T_r \right)(\sigma_b - E_r) \Big] .
\label{si:growth_rate_het}
\end{align}

Equation \eqref{si:growth_rate_het} reduces correctly in two limits. Setting
$E_l = E_r = E$, $Z_l = Z_r = Z$ and $L_l = L_r = L/2$ returns
$2(\sigma_b - E)Z^{-1}\tanh(L/2\ell)$, the homogeneous result of the main text.
In the limit $L_\alpha \gg \ell_\alpha$, $S_\alpha \to 0$ and
$T_\alpha \to 1$, so that \eqref{si:matrix} collapses to
\begin{align}
\dot{x}_l = -w_l, \qquad \dot{x}_I = \Delta, \qquad \dot{x}_r = w_r ,
\label{si:asymptotic_velocities}
\end{align}
with $w_\alpha = (\sigma_b-E_\alpha)/Z_\alpha$ and
$\Delta = (E_r-E_l)/(Z_l+Z_r)$, and $dL^\ast/dt \to w_l + w_r$. Each edge
advances at exactly the speed a homogeneous tissue made of that part would have:
once the tissue is much longer than the hydrodynamic lengths the two edges are
screened from one another, and only the interface retains any memory of the
heterogeneity.

\section{Frozen fractions versus proportionate growth}
\label{si:proportionate}

\subsection{The distinction}

From \eqref{si:asymptotic_velocities} the two parts grow at rates
$\dot{L}_l = w_l + \Delta$ and $\dot{L}_r = w_r - \Delta$, so their length
fractions approach the constants
\begin{align}
\frac{L_l}{L} = \frac{\dot{L}_l}{\dot{L}^\ast}
= \frac{Z_r}{Z_l+Z_r} + \frac{2\Delta}{\dot{L}^\ast} ,
\qquad
\frac{L_r}{L} = \frac{Z_l}{Z_l+Z_r} - \frac{2\Delta}{\dot{L}^\ast} .
\label{si:fractions}
\end{align}
The fractions are \emph{frozen}, in that they stop changing at long times, but
they are not in general \emph{preserved}: the values at which they freeze are
determined entirely by the material parameters $E_\alpha, \eta_\alpha,
\gamma_\alpha$ and the applied stress, and bear no relation to the proportions
$L_{\alpha 0}/L_0$ the tissue started with. A tissue prepared with a short stiff
part and a long soft one ends with quite different proportions, reached
asymptotically and then held fixed.

Proportionate growth in the biological sense is the stronger statement that the
proportions present at the outset are the ones maintained. Within this model it
is a condition on the parameters rather than a generic property of the
perennially growing phase.

\subsection{The condition for proportion preservation}

Write $\varphi = L_{l0}/L_0$ for the initial fraction occupied by the left part.
Demanding $L_l/L = \varphi$ in \eqref{si:fractions} gives, after rearrangement,
\begin{align}
\Delta = \varphi \, w_r - (1-\varphi) \, w_l ,
\label{si:proportionate_condition}
\end{align}
that is,
\begin{align}
\frac{E_r-E_l}{Z_l+Z_r}
= \varphi \, \frac{\sigma_b-E_r}{Z_r} - (1-\varphi) \, \frac{\sigma_b-E_l}{Z_l} .
\label{si:proportionate_explicit}
\end{align}
This is a \emph{single} scalar constraint on the six material parameters
$E_\alpha, \eta_\alpha, \gamma_\alpha$ (entering through $E_\alpha$ and
$Z_\alpha = \sqrt{\eta_\alpha\gamma_\alpha}$) and the applied stress $\sigma_b$.
There is accordingly a five-parameter family of tissues, all with different
material properties, that grow while preserving any prescribed set of initial
proportions. This is the redundancy referred to in the main text: the emergent
growth rate and the emergent proportions depend on the material parameters only
through a small number of combinations, so that the parameters may be varied
along the level sets of \eqref{si:proportionate_explicit} without disturbing the
proportions.

\subsection{A transparent special case}

The condition simplifies considerably when the two parts share an elastic
modulus. Setting $E_l = E_r$ makes the interface stationary relative to the
material, $\Delta = 0$, and \eqref{si:fractions} reduces to
$L_l/L = Z_r/(Z_l+Z_r)$: the proportions are set entirely by the ratio of
impedances. Preservation of the initial proportions then requires
\begin{align}
\frac{Z_l}{Z_r} = \frac{L_{r0}}{L_{l0}} ,
\label{si:special_case}
\end{align}
that is, the impedances must be in inverse proportion to the initial lengths.
An edge advances at speed $w_\alpha = (\sigma_b-E_\alpha)/Z_\alpha$, so a part
with larger impedance grows more slowly, and for the two parts to keep pace in
relative terms the more heavily damped part must be the initially shorter one.
Conversely, if the two parts have equal impedances, \eqref{si:fractions} shows
that equal halves are maintained only if $E_l = E_r$, that is, only if the
tissue is homogeneous.

\subsection{Choosing parameters for a proportion-preserving simulation}
\label{si:tuning}

The numerics of Fig.~3 of the main text integrate the full model, with active
stresses generated in the bulk and the density evolving dynamically. The
mapping from the imposed-stress problem is
$\sigma_b \to \zeta f(\rho_0)$, so the results above apply with the edge speeds
\begin{align}
w_\alpha = \frac{\zeta f(\rho_0) - E_\alpha}{Z_\alpha} .
\end{align}
We take the activity $\zeta$ and the setpoint $\rho_0$ to be common to the two
parts, so that the heterogeneity resides entirely in the mechanical parameters.

\paragraph*{Equal elastic moduli.} The simplest tuning takes
$E_l = E_r = E$. The interface drift then vanishes, $\Delta = 0$, both edges
advance at speeds differing only through the impedances, and \eqref{si:fractions}
reduces to $L_l/L = Z_r/(Z_l+Z_r)$. Imposing $L_l/L = L_{l0}/L_0$ gives
\begin{align}
\frac{Z_l}{Z_r} = \frac{L_{r0}}{L_{l0}}
\qquad \mbox{i.e.} \qquad
\frac{\eta_l \gamma_l}{\eta_r \gamma_r} = \left( \frac{L_{r0}}{L_{l0}} \right)^{2} .
\label{si:tuning_condition}
\end{align}
The initially shorter part must be the more heavily damped one, so that its
edge advances more slowly by exactly the factor needed to keep the two parts in
step. Note that the condition involves the product $\eta_\alpha\gamma_\alpha$
only, and is independent of $\zeta$, $f(\rho_0)$ and $E$: the same pair of
impedances preserves the proportions at any value of the control parameter, so
long as the tissue is in the perennially growing phase.

\paragraph*{How to realise the impedance ratio.} Equation
\eqref{si:tuning_condition} fixes only $Z_l/Z_r$ and leaves the hydrodynamic
lengths $\ell_\alpha = \sqrt{\eta_\alpha/\gamma_\alpha}$ free. It is convenient
to scale the viscosity and the friction of the left part together,
\begin{align}
\frac{\eta_l}{\eta_r} = \frac{\gamma_l}{\gamma_r} = a ,
\qquad
a = \frac{L_{r0}}{L_{l0}} ,
\label{si:tuning_scaling}
\end{align}
which gives $Z_l/Z_r = a$ as required while leaving $\ell_l = \ell_r \equiv \ell$
unchanged. Equal hydrodynamic lengths are convenient because the asymptotic
analysis holds only for $L_\alpha \gg \ell_\alpha$: with
\eqref{si:tuning_scaling} a single condition $L_{\alpha 0} \gg \ell$ controls the
approach to the asymptotic regime in both parts. Tuning the impedance ratio
through the viscosities alone would lengthen the screening length of the
initially shorter part, where that condition is hardest to satisfy.

\paragraph*{Requirements from the density dynamics.} Two further conditions come
from the fact that Fig.~3 solves the full problem rather than the
imposed-stress reduction.

First, the results of this section neglect the density-gradient contribution to
the edge velocity. From Sec.~\ref{si:greens} this contribution is bounded by
$\zeta \gamma_\alpha^{-1}\max_{t,x}\vert\partial_x f\vert$, so it is negligible
compared with $w_\alpha$ only when the birth--death rate is fast enough to hold
$\rho$ near $\rho_0$ throughout the tissue. In terms of the dimensionless rate
used in the phase diagram this requires
\begin{align}
\kappa \tau_\alpha \gg 1 , \qquad \tau_\alpha = \frac{\eta_\alpha}{E} ,
\label{si:kappa_condition}
\end{align}
in both parts. With the scaling \eqref{si:tuning_scaling}, $\tau_l/\tau_r = a$,
so the part with the larger $\eta$ sets the requirement and $\kappa$ must be
chosen against $\tau_l$.

Second, the correction scales as $\gamma_\alpha^{-1}$ and the two parts have
different frictions, so the residual density-gradient effect is asymmetric: it
is larger in the part with the smaller $\gamma$. This produces a small,
systematic drift of the fractions away from the tuned values, which decreases as
$\kappa$ is increased. The same is true of the small diffusivity $D$ added to
the density equation for numerical stability, which introduces a further length
$\sqrt{D/\kappa}$ that must be kept small compared with $\ell$ and with the
part lengths.

\paragraph*{Summary.} A proportion-preserving run may be set up by choosing
$E_l = E_r$; $\eta_l/\eta_r = \gamma_l/\gamma_r = L_{r0}/L_{l0}$; $\zeta f(\rho_0) > E$
so that the tissue is in the perennially growing phase; $\kappa\tau_l \gg 1$;
and $L_{\alpha 0} \gg \ell$. The predicted result is a pair of horizontal lines
at $L_{l0}/L_0$ and $L_{r0}/L_0$, in contrast with the drifting curves obtained
for untuned parameters.

\subsection{Validity}

For \eqref{si:fractions} to describe a physically sensible partition both
fractions must be positive, which requires the interface to outrun neither edge,
\begin{align}
-w_l < \Delta < w_r .
\label{si:positivity}
\end{align}
When this is violated the interface overtakes one of the edges, the softer part
progressively consumes the stiffer one, and the composite tissue does not
maintain any fixed proportions. Since \eqref{si:asymptotic_velocities} holds
only for $L_\alpha \gg \ell_\alpha$, all of the statements of this section are
asymptotic; at early times, while the tissue is comparable in size to the
hydrodynamic lengths, the fractions still drift.

\section{Isotropic steady state in \texorpdfstring{$d$}{d} dimensions}
\label{si:ddim}

We derive here the steady-state size of an isolated, isotropic tissue in $d$
spatial dimensions with active stresses in the bulk.

\subsection{Setting up}

In the steady state $\vect{v} = \vect{0}$, so the viscous stress vanishes
identically and the substrate friction $\vect{F}_{\mathrm{ext}} = -\gamma\vect{v}$
does no work; force balance reduces to $\nabla\cdot\tens{\sigma} = 0$ with
$\tens{\sigma} = \tens{\sigma}_{\mathrm{elastic}} + \tens{\sigma}_{\mathrm{active}}$.
The density equation $D_t\rho = -(\nabla\cdot\vect{v})\rho + \kappa\rho(1-\rho/\rho_0)$
has the two fixed points $\rho = 0$ and $\rho = \rho_0$; taking the latter, the
active stress is uniform,
\begin{align}
\tens{\sigma}_{\mathrm{active}} = -\zeta \, f(\rho_0) \, \tens{I} .
\end{align}
The tissue is isolated, so the traction vanishes on the boundary,
$\vect{\hat{n}}\cdot\tens{\sigma}\vert_{\partial\Omega_t} = \vect{0}$.

\subsection{The dilation ansatz}

Isotropy suggests looking for a uniform dilation about the centre,
\begin{align}
\vect{u}^\star = c \, \vect{x} ,
\label{si:dilation}
\end{align}
with $c$ a constant to be determined. Then $\nabla\vect{u} = c\,\tens{I}$, so
the strain is $\tens{\varepsilon} = c\,\tens{I}$ with
$\mathrm{tr}\,\tens{\varepsilon} = d\,c$. The deviatoric part vanishes,
\begin{align}
\tens{\varepsilon} - \frac{1}{d}\left( \mathrm{tr}\,\tens{\varepsilon} \right) \tens{I}
= c\,\tens{I} - c\,\tens{I} = 0 ,
\end{align}
so that the shear modulus $K$ drops out of the problem entirely and only the
bulk modulus $E$ appears:
\begin{align}
\tens{\sigma}_{\mathrm{elastic}} = E \left( \mathrm{tr}\,\tens{\varepsilon} \right) \tens{I}
= E \, d \, c \, \tens{I} .
\end{align}
The total stress
$\tens{\sigma} = \left[ E d c - \zeta f(\rho_0) \right] \tens{I}$
is spatially uniform, so $\nabla\cdot\tens{\sigma} = 0$ is satisfied
identically and \eqref{si:dilation} is consistent with force balance in the
bulk for any $c$. The constant is fixed instead by the boundary condition:
$\vect{\hat{n}}\cdot\tens{\sigma} = \left[ E d c - \zeta f(\rho_0)\right]\vect{\hat{n}}$
vanishes only if
\begin{align}
c = \frac{\zeta \, f(\rho_0)}{d \, E} .
\label{si:c}
\end{align}

\subsection{From strain to size}

Since $\vect{u} = \vect{x} - \vect{X}$, \eqref{si:dilation} gives
$\vect{X} = (1-c)\,\vect{x}$, i.e. $\vect{x} = \vect{X}/(1-c)$: every material
point moves radially outward by the same factor, so the tissue changes size
without changing shape. Any linear dimension is therefore multiplied by
$(1-c)^{-1}$, and we may take $R$ below to denote any such dimension -- the
radius, if the tissue is a $d$-dimensional ball, but equally the diameter or the
edge of a cube. Equation \eqref{si:c} was obtained from a stress that is uniform
and isotropic, so the traction $\vect{\hat{n}}\cdot\tens{\sigma}$ vanishes on a
boundary of any shape; the result does not assume a spherical tissue. With
\eqref{si:c},
\begin{align}
\frac{R^\star}{R_0} = \left( 1 - \frac{\zeta f(\rho_0)}{d\,E} \right)^{-1} ,
\qquad
\frac{V^\star}{V_0} = \left( 1 - \frac{\zeta f(\rho_0)}{d\,E} \right)^{-d} ,
\label{si:ddim_size}
\end{align}
the second following from $\Lambda = \det\tens{J} = (1-c)^{-d}$. The associated
fields are
\begin{align}
\vect{u}^\star = \frac{\zeta f(\rho_0)}{d\,E} \, \vect{x} ,
\qquad
\rho^\star = \rho_0 .
\end{align}

\subsection{Remarks}

Setting $d=1$ in \eqref{si:ddim_size} returns
$L^\star/L_0 = (1-\zeta f(\rho_0)/E)^{-1}$, the one-dimensional result quoted in
the main text, which in turn is the imposed-stress result with
$\sigma_b \to \zeta f(\rho_0)$.

The growth factor \eqref{si:ddim_size} diverges, and the steady state ceases to
exist, when
\begin{align}
\zeta f(\rho_0) = d \, E .
\label{si:ddim_threshold}
\end{align}
The threshold is therefore $d$-dependent: a tissue in higher dimensions
tolerates a proportionally larger active stress before growing without bound.
The reason is that given in Sec.~\ref{si:constitutive}: what is bounded is the
strain, $c<1$, and the isotropic elastic stress that this
strain can generate is $E\,d\,c < d\,E$, the factor $d$ arising because a
dilation of magnitude $c$ in each of $d$ directions contributes $d\,c$ to
$\mathrm{tr}\,\tens{\varepsilon}$. The threshold $\zeta f(\rho_0) = E$ quoted in
the main text is thus specific to $d=1$.

The shear modulus $K$ is absent from every expression above, a consequence of
isotropy alone: a purely dilational deformation produces no deviatoric strain,
so a tissue whose growth is isotropic cannot be used to infer $K$.

\section{Green's function for the active problem}
\label{si:greens}

In the model with active stresses the one-dimensional force balance reads
\begin{align}
\gamma v = E \, \partial_x^2 u + \eta \, \partial_x^2 v - \zeta \, \partial_x f(\rho) .
\end{align}
In the perennially growing phase $u^\ast = x$, so $\partial_x^2 u = 0$ and the
velocity obeys the inhomogeneous modified Helmholtz equation
\begin{align}
\left( \gamma - \eta \, \partial_x^2 \right) v = -\zeta \, \partial_x f(\rho) .
\label{si:helmholtz}
\end{align}
Let $G(x,y)$ be the Green's function of the operator on the left, that is
\begin{align}
\left( \gamma - \eta\,\partial_x^2 \right) G(x,y) = \delta(x-y) ,
\qquad
\partial_x G \big\vert_{x = \pm L^\ast/2} = 0 .
\end{align}
Building it from the homogeneous solutions $\cosh\left[(x+L^\ast/2)/\ell\right]$
and $\cosh\left[(x-L^\ast/2)/\ell\right]$, which satisfy the left and right
boundary conditions respectively, and matching the unit jump in
$-\eta \, \partial_x G$ across $x=y$, gives
\begin{align}
G(x,y) = \frac{\ell}{\eta} \,
\frac{\cosh\left( \dfrac{x_< + L^\ast/2}{\ell} \right)
      \cosh\left( \dfrac{x_> - L^\ast/2}{\ell} \right)}
     {\sinh\left( \dfrac{L^\ast}{\ell} \right)} ,
\label{si:greens_function}
\end{align}
where $x_< = \min(x,y)$ and $x_> = \max(x,y)$, and $\ell = \sqrt{\eta/\gamma}$.

Two properties of \eqref{si:greens_function} are used in the main text. First,
$G$ does not change sign: $\cosh$ is positive for any real argument and
$\sinh(L^\ast/\ell) > 0$ for $L^\ast>0$, so that $G(x,y) > 0$ for every $x$ and
$y$. Second, integrating over the source position,
\begin{align}
\int_{-L^\ast/2}^{L^\ast/2} G(x,y) \, dy = \frac{1}{\gamma} ,
\label{si:greens_integral}
\end{align}
independently of $x$, the response to a uniform unit forcing being the uniform
velocity $1/\gamma$, for which the viscous term vanishes identically.

Together these bound the density-gradient contribution to the velocity. Since
$G$ is of one sign, $\int |G| \, dy = \int G \, dy = \gamma^{-1}$, and therefore
\begin{align}
\left| \zeta \int \frac{\partial f}{\partial y} \, G(x,y) \, dy \right|
\leq \frac{\zeta}{\gamma} \, \max_{t,x} \left| \partial_x f \right| .
\end{align}
The positivity of $G$ is what permits $\int|G|\,dy$ to be replaced by
$\int G \, dy$ here; had $G$ changed sign, \eqref{si:greens_integral} would not
have supplied the bound. The correction is negligible when the birth--death rate
$\kappa$ is large enough to hold the density close to its setpoint $\rho_0$.

\section{Numerical method}
\label{si:numerics}

All numerical solutions reported in the main text were obtained with a
finite-element discretisation implemented in FEniCS \cite{si:fenics}. The
scheme is an updated-Lagrangian one \cite{si:belytschko}.

\subsection{Regularisation of the density equation}

The active-stress regulation function used throughout is
\begin{align}
f(\rho) = \frac{\rho \left( \hat\rho - \rho \right)}{\rho^2 + \rho_s^2} ,
\label{si:f_form}
\end{align}
whose form is dictated by the sign the active stress must take at each density.
Division at low density pushes material outward while extrusion and apoptosis at
high density draw it inward, so the active stress should be extensile for
$\rho < \hat\rho$ and contractile for $\rho > \hat\rho$. The density $\hat\rho$
is thus a homeostatic set point at which division and apoptosis balance and the
active stress changes sign, $f(\hat\rho) = 0$. The factor $\rho$ in the numerator
ensures that no active stress is generated where there is no tissue, and the
denominator keeps $f$ bounded, so that the active stress saturates at high
density rather than growing without limit, with $\rho_s$ setting the density at
which it does so. The mechanical set point $\hat\rho$ and the logistic carrying
capacity $\rho_0$ are independent parameters, and it is precisely because they
differ that the tissue carries an active stress $\zeta f(\rho_0) \neq 0$ in the
steady state.

A small diffusivity $D$ is added to the density equation
\eqref{si:density_reg} for numerical stability,
\begin{align}
D_t \rho = -(\nabla\cdot\vect{v})\,\rho
+ \kappa\rho\left(1-\frac{\rho}{\rho_0}\right)
+ D \nabla^2 \rho .
\label{si:density_reg}
\end{align}
The diffusion term requires a boundary condition on $\rho$, which the
undiffused problem does not. We have verified that taking either
$\rho\vert_{\partial\Omega_t} = \rho_0$ or $\rho\vert_{\partial\Omega_t} = 0$
leaves the results unchanged, and that the particular form chosen for $f(\rho)$
affects them only quantitatively: neither the location of the transition at
$\zeta f(\rho_0)/E = 1$ nor the asymptotic growth rate depends on these choices.
As noted in Sec.~\ref{si:tuning}, $D$ introduces a further length
$\sqrt{D/\kappa}$, which must be kept small compared with the hydrodynamic
length $\ell$ and with the size of the tissue.

\subsection{The deforming mesh}

The domain is not fixed: the tissue occupies $\Omega_t$, whose extent is itself
part of the solution. One may map the problem onto a reference domain of fixed
size, as is done analytically in Sec.~\ref{si:stability}, at the cost of
introducing advective terms and time-dependent coefficients. We instead deform
the computational mesh with the material and solve the equations on the
deforming mesh directly.

The mesh nodes are therefore treated as material points. Within each time step
the mesh is held fixed and the equations for $u$, $v$ and $\rho$ are solved on
the current configuration, so that all spatial derivatives are evaluated on the current domain. At the end
of the step the nodes are displaced according to $\dot{\vect{x}} = \vect{v}$,
which advances $\Omega_t$ and supplies the geometry for the next step. The
convective part of the material derivative $D_t$ is thus not discretised explicitly but is accounted for
by the motion of the mesh itself, and the boundary $\partial\Omega_t$ is tracked
exactly rather than captured on a fixed grid. This is the numerical counterpart
of the statement in the main text that the kinematics and the constitutive law
are not separable in this model.

Because the nodes follow the material, the elements stretch as the tissue grows,
without bound in the perennially growing phase, and the domain is therefore
remeshed periodically. Once the elements have become sufficiently distorted the
current domain is remeshed and the fields $u$, $v$ and $\rho$ are
transferred onto the new mesh by projection. We used first-order Lagrange elements, forward Euler scheme with $\Delta t/\tau = 10^{-2}$, the mesh resolution was always such that $\Delta x / \ell < 5\times10^{-2}$. For the model with the density field, we added a small diffusivity $(\tau/\ell^2) D = 10^{-1}$.

\section{Description of the supplementary movies}
\label{si:movies}

The movies below show numerical solutions of the full one-dimensional model,
that is equations (1), (2) and (3) of the main text with total stress
$\sigma = E\partial_x u + \eta\partial_x v - \zeta f(\rho)$, together with the
density equation and the traction-free condition $\sigma\vert_{x=\pm L(t)/2}=0$.
A small diffusivity is added to the density equation for numerical stability, as
described in the main text. In each movie the tissue is drawn on its current
domain $\Omega_t = (-L(t)/2, L(t)/2)$, so that the domain itself is seen to
deform; the displacement $u$, velocity $v$ and cell number density $\rho$ are
plotted as functions of position at each instant.

\begin{itemize}

\item \textbf{Movie S1 -- Approach to the determinate steady state.}
Spatiotemporal evolution of $u$, $v$ and $\rho$ for
$\zeta f(\rho_0)/E < 1$. The tissue expands from its initial configuration and
comes to rest: the velocity decays to zero everywhere, the displacement settles
to the linear profile $u^\star = \left[\zeta f(\rho_0)/E\right] x$, the density
relaxes to the setpoint $\rho_0$, and the domain length saturates at the value
given by \eqref{si:ddim_size} with $d=1$. This is the determinate phenotype.
$\zeta f(\rho_0)/E = 0.5$, $\kappa\tau = 1$.

\item \textbf{Movie S2 -- Perennially growing state.}
The same fields for $\zeta f(\rho_0)/E > 1$. The domain now grows without bound
and asymptotically linearly in time. Note that the velocity vanishes at the
midpoint and is largest at the two edges, and that as the tissue lengthens the
region of non-zero strain rate becomes confined to within a hydrodynamic length
$\ell$ of each boundary: the interior becomes quiescent and the growth is
carried entirely by the periphery, even though cell division and death continue
throughout the bulk. The density remains close to $\rho_0$ except in these
boundary layers. This is the indeterminate phenotype.
$\zeta f(\rho_0)/E = 2.5$, $\kappa\tau = 1$.

\item \textbf{Movie S3 -- Tissue-scale oscillations at low birth--death rate.}
For sufficiently small $\kappa\tau$ the tissue does not settle into either of
the two special solutions but instead oscillates as a whole, the length and the
density alternately overshooting and undershooting their steady values. The
mechanism is the delay between a change in density and the active stress it
generates: with the contractile--extensile form of $f(\rho)$ used here, a
region that becomes too dense generates a contractile stress that expels
material, which then overshoots, and the slow logistic relaxation at small
$\kappa$ is unable to damp the excursion within one mechanical relaxation time
$\tau$. The oscillations disappear as $\kappa\tau$ is increased. $\zeta f(\rho_0)/E = 0.5$, $\kappa\tau = 0.1$.

\end{itemize}



\end{document}